\documentclass[twocolumn, switch]{article} 

\usepackage{preprint}

\usepackage{amsmath, amsthm, amssymb, amsfonts}

\usepackage[sort&compress,numbers,square]{natbib}
\usepackage[utf8]{inputenc}	
\usepackage[T1]{fontenc}	
\usepackage{xcolor}		
\usepackage[colorlinks = true,
            linkcolor = purple,
            urlcolor  = blue,
            citecolor = cyan,
            anchorcolor = black]{hyperref}	
\usepackage{booktabs} 		
\usepackage{nicefrac}		
\usepackage{microtype}		
\usepackage{lineno}		
\usepackage{float}			

\usepackage{newfloat}
\DeclareFloatingEnvironment[name={Supplementary Figure}]{suppfigure}
\usepackage{sidecap}
\sidecaptionvpos{figure}{c}

\usepackage{titlesec}
\titlespacing\section{0pt}{12pt plus 3pt minus 3pt}{1pt plus 1pt minus 1pt}
\titlespacing\subsection{0pt}{10pt plus 3pt minus 3pt}{1pt plus 1pt minus 1pt}
\titlespacing\subsubsection{0pt}{8pt plus 3pt minus 3pt}{1pt plus 1pt minus 1pt}

\usepackage{hyperref}
\usepackage{algorithm} 
\usepackage{algpseudocode} 
\usepackage{doi}

\title{Bell Inequalities from Polyhedral Sampling}

\usepackage{authblk}

\author[1]{Christian Staufenbiel}

\begin{document}

\twocolumn[
  \begin{@twocolumnfalse} 

\maketitle

\begin{abstract}
Bell inequalities play a central role in certifying quantum correlations and underpin protocols such as device-independent quantum key distribution. However, enumerating all Bell inequalities for a given scenario remains intractable beyond the simplest cases, as it requires solving a computationally hard facet enumeration problem on the associated Bell polytope. We propose the Adjacency Sampling method, which builds on the Adjacency Decomposition method but sacrifices completeness for speed. On previously solved Bell polytopes, the method reproduces every known class of inequalities. For scenarios where no complete enumeration exists, it greatly exceeds existing partial results: in $\mathcal{L}_{3,3,3,3}$ we obtain over $1.29 \times 10^8$ classes, more than 25 times the previous count; in $\mathcal{L}_{4,5,2,2}$ we nearly triple the known list to 49\,358 classes; and for $\mathcal{L}_{4,6,2,2}$ we report over 4.3 million classes.
\end{abstract}

\vspace{0.35cm}

  \end{@twocolumnfalse} 
] 


\section{Introduction}

Consider two spatially separated parties, each choosing from a set of measurement settings and observing one of several possible outcomes. This setup, known as a Bell scenario, lies at the heart of the study of non-locality in quantum mechanics. Bell showed that the correlations arising between the two parties from any local classical theory must satisfy certain linear inequalities~\cite{Bell1964}, now known as Bell inequalities. A violation of such an inequality certifies that the observed correlations are genuinely quantum and cannot be explained by any local classical model. Such quantum correlations have been confirmed in a loophole-free experimental test~\cite{Hensen2015}. Beyond their foundational role, Bell inequalities have found practical applications, most notably in device-independent quantum key distribution (DIQKD)~\cite{Acin2007}, where the security of protocols relies on the verified violation of such inequalities.

A Bell scenario is characterized by the number of settings and outcomes available to each party. Determining all Bell inequalities for a given scenario amounts to finding the facets of the associated Bell polytope~\cite{Brunner2014}, a problem that is computationally hard in general~\cite{Avis1997}. While several exact algorithms exist~\cite{Fukuda1996, Avis1996, Christof2001}, they quickly become intractable as the number of settings and outcomes grows. For this reason, the complete set of Bell inequalities is known only for a few small scenarios.

Recently, alternative approaches have been proposed that aim to generate possibly incomplete lists of Bell inequalities~\cite{Jesus2023, Cope2019}. In this paper, we introduce the \textit{Adjacency Sampling (AS) method}, a modification of the Adjacency Decomposition Method~\cite{Christof2001} that sacrifices completeness for a drastically reduced runtime. Although designed as a sampling technique, it recovers all known Bell inequalities, up to symmetry, for all tested, previously enumerated scenarios. To further test the method's capabilities, we also validate it against enumerated Cut polytopes.

Applying the method to larger, previously unsolved scenarios yields a significant number of new Bell inequalities. For $\mathcal{L}_{3,3,3,3}$, we identify more than $1.29 \times 10^8$ classes of inequalities, compared to the previously known 4.8 million~\cite{Liu2024}. For $\mathcal{L}_{4,5,2,2}$, we find 49\,358 classes, extending the 18\,277 reported by Cope and Colbeck~\cite{Cope2019}. We also present the first results for $\mathcal{L}_{4,6,2,2}$, identifying more than 4.3 million classes of Bell inequalities.

The remainder of this paper is organized as follows. Section~2 introduces the necessary background on convex polytopes, the dual description problem, and Bell and Cut polytopes. Section~3 reviews the Adjacency Decomposition Method. Section~4 presents the Adjacency Sampling Method and its implementation. Section~5 evaluates the method and presents new Bell inequalities. Section~6 concludes and gives an outlook for future research.

\section{Convex Polytopes}
This section introduces the concepts for polytopes required in this work. For more thorough discussions on polytopes and their features refer to the books by Grünbaum~\cite{Grnbaum2003} or Ziegler~\cite{Ziegler1995}.

A convex polytope $\mathcal{P}$ is defined as the \textit{convex hull} over a finite set $\mathcal{K} = \{ x_1, \ldots, x_n \}$, where $x_i \in \mathbb{R}^d$:

\begin{equation}
    \label{eq:v_representation}
    \mathcal{P} = conv(\mathcal{K}) = \{ \lambda_1 x_1 + \ldots + \lambda_n x_n \; : \; \lambda_i \geq 0, \; \sum_i \lambda_i = 1\} 
\end{equation}
The minimal representation of $\mathcal{P}$ is uniquely identified by its extremal points / vertices $\mathcal{V}(\mathcal{P})$ (V-representation).

Equivalently, $\mathcal{P}$ can be described as the intersection of $m$ half-spaces (H-representation) 
\begin{equation}
    \label{eq:h_representation}
    \mathcal{P} = \{ x \in \mathbb{R}^d \; | \; A \cdot x \leq b \},
\end{equation}
where $A \in \mathbb{R}^{m \times d}$ and $b \in \mathbb{R}^m$.

Any non-empty intersection of a supporting hyperplane $\mathcal{H}$ in $\mathbb{R}^d$ (i.e. $\mathcal{H}$ does not cut through $\mathcal{P}$) is called a face $\mathcal{F}$ of $\mathcal{P}$ 
\begin{equation}
	\mathcal{F} = \mathcal{H} \cap \mathcal{P}.
	\label{eq:face_definition}
\end{equation}
It is important to note that $\mathcal{F}$ is again a polytope of dimension $k \leq dim(\mathcal{P})$, denoted as $k$-face or $k$-polytope.
For a $d$-polytope $\mathcal{P}$, the $(d-1)$-faces are called facets, and the description of $\mathcal{P}$ by the facet-defining inequalities is the minimal H-representation.

\subsection{Automorphisms}
The combinatorial automorphism group of a polytope $Aut(\mathcal{P})$ contains all automorphisms of $\mathcal{P}$ that preserve the face-inclusion relation, i.e. any face $\mathcal{F}$ remains a face of $\mathcal{P}$ under the transformation.

Such transformations of $\mathcal{P}$ can be represented by permutations $\pi \in S_n$ of its vertices:

\begin{equation*}
	\pi(\mathcal{P}) = conv(\{ v_{\pi(1)}, \ldots, v_{\pi(n)}\}) = \mathcal{P}.
\end{equation*}
The automorphism group $Aut(\mathcal{P})$ is a subgroup of the symmetric group $S_n$ where $n = |\mathcal{V}(\mathcal{P})|$.

The automorphism group induces an equivalence relation on the faces of a polytope, e.g. for two faces $\mathcal{F}_1, \mathcal{F}_2$: 
\begin{equation*}
	\mathcal{F}_1 \equiv \mathcal{F}_2 \leftrightarrow \exists \, \pi \in Aut(\mathcal{P}) \; with \; \mathcal{F}_1 = \pi(\mathcal{F}_2)
\end{equation*}
A set of equivalent facets is called an \textit{orbit} or \textit{facet-class}.

\subsection{Dual Description Problem}
Transforming between the H- and V-representation of any polytope is known as the dual description problem. \textit{Vertex Enumeration} describes the process of obtaining the vertices $\mathcal{V}(\mathcal{P})$ from the H-representation, while the opposite process is known as \textit{Facet Enumeration}.

The conversion has wide-ranging applications in combinatorial optimization~\cite{Deza1997} and quantum information~\cite{Brunner2014}. Several general-purpose algorithms have been developed for this task: Double Description Method~\cite{Fukuda1996}, Reverse search algorithms~\cite{Avis1996}, and beneath-and-beyond algorithms~\cite{Joswig2002}, each subject to limitations in terms of scalability or suitability for specific polytope structures. In particular, the exponential behavior of the problem leads to extensive runtimes with increasing size of the input~\cite{Avis1997}.
Recently, more efficient implementations of known methods yielded complete H-representations of polytopes, which were previously unavailable~\cite{Deza2015}.

\subsection{Bell Polytopes}
Originally, Bell inequalities provided mathematical constraints to test if reality obeys \textit{local realism}, a physical principle violated by the formulation of quantum mechanics \cite{EPR1935, Bell1964}.

Besides ruling out local-hidden-variable theories~\cite{Hensen2015}, Bell inequalities found various applications such as Quantum Key Distribution~\cite{Zapatero2023}.

Bell inequalities are the facets of the so-called 
 \textit{Bell Polytope} or \textit{Local Polytope} $\mathcal{L}$. The task of finding Bell inequalities is thus a facet enumeration problem on such polytopes. A detailed introduction to the Bell Polytope can be found elsewhere~\cite{Brunner2014}.

The vertices $\mathcal{V}(\mathcal{L})$ can simply be derived from all possible deterministic correlations between the parties in a Bell experiment. In this work, we consider such experiments with two parties, where the first (second) party has $m_a$ ($m_b$) inputs and $n_a$ ($n_b$) outputs. The corresponding polytope is referred to by $\mathcal{L}_{m_a,m_b,n_a,n_b}$. The Bell scenario yields symmetries due to relabelling of inputs, outputs and parties, which determine the automorphism group of the polytope.

\subsection{Cut Polytopes} 

A cut over a graph $\mathcal{G}$ is the partition of the graph into two subsets of vertices. The cut polytope $\textit{CUT}(\mathcal{G})$ is generated by all possible cuts through the graph. These polytopes are tightly connected to NP-hard problems such as the \textit{max-cut} problem or the unconstrained quadratic 0,1 programming problem~\cite{Deza1997}.

In this work we use cut polytopes over \textit{complete (multipartite) graphs} $K_{i,j,k,\ldots}$ to benchmark the introduced sampling method. The explicit construction of the polytope can be found elsewhere~\cite{Deza1997}. Note that the cut polytopes over the graphs $K_{1,j,k}$ are equivalent to the bipartite binary-outcome Bell polytopes $\mathcal{L}_{j,k,2,2}$.

Recently, the Adjacency Decomposition method, introduced in the next section, was applied to several cut polytopes, yielding previously unknown H-representations and Bell inequalities~\cite{Deza2015}.

\section{Adjacency Decomposition Method}

The \textit{Adjacency Decomposition (AD) Method} is a facet-enumeration method that traverses along the surface (facets) of a polytope $\mathcal{P}$~\cite{Christof2001}.

Given the V-representation of a polytope, an initial facet can be found efficiently using linear programming~\cite{Cope2019}. We denote this operation by $\mathcal{F} = LP(\mathcal{P})$. Beginning with an initial facet $\mathcal{F}$, the method uses a \textit{rotation algorithm} to generate the adjacent facets of $\mathcal{F}$. Applying the rotation algorithm again to the newly found facets, the method enumerates the complete surface of $\mathcal{P}$ sequentially. The AD method is summarized in Alg.~\ref{alg:adm}. 

The rotation algorithm rotates a facet $\mathcal{F}$ around $(d-2)$-face $\mathcal{H}$ and generates a new facet $\mathcal{F}_{new}$. The two facets then share the ridge $\mathcal{H}$. We denote such a rotation by $\mathcal{F}_{new} = rotate(\mathcal{F}, \mathcal{H})$. The details of this algorithm are explained in~\cite{Loerwald2015}. 

\subsection{Symmetries}

In the symmetric variant, a newly found facet $\mathcal{G}$ is only explored further if no equivalent facet has been previously considered. The output of the algorithm is reduced to the facet-classes of $\mathcal{P}$. The number of facet-classes is often much smaller than the number of facets~\cite{Deza2015} and the runtime of the AD method is reduced when exploiting its symmetry properties.

The naive approach to verify the equivalence of two facets $\mathcal{F}_1,\mathcal{F}_2$ is to iteratively apply all elements of $Aut(\mathcal{P})$ and check for equality. A more efficient method is partition-backtracking based on stabilizer chains~\cite{Leon1991}, as implemented by GAP~\cite{GAP4}.

\begin{algorithm}
\caption{Symmetric AD method.}
\label{alg:adm}
\begin{algorithmic}
\State \textbf{Input: } $\mathcal{V}(\mathcal{P})$
\State \textbf{Output: } All facet-classes of $\mathcal{P}$
\State $\mathcal{F}_0 = LP(\mathcal{P})$
\State $\mathbf{C} := \{ \mathcal{F}_0 \}$ \Comment{Facets to be considered}
\State $\mathbf{O} := \{ \mathcal{F}_0 \}$ \Comment{All found facets}
\For{$\mathcal{F} \in \mathbf{C}$}
\State $\mathbf{H} := \text{all facets of} \; \mathcal{F}$ \Comment{Possible recursion}
\For{$\mathcal{H} \in \mathbf{H}$}
\State $\mathcal{F}_{new} = rotate(\mathcal{F}, \mathcal{H})$
\If{$\mathcal{F}_{new}$ inequivalent to all facets in $\mathbf{O}$}
\State $\mathbf{O} = \mathbf{O} \cup \{ \mathcal{F}_{new} \}$
\State $\mathbf{C} = \mathbf{C} \cup \{ \mathcal{F}_{new} \}$
\EndIf
\EndFor
\EndFor
\State \textbf{return} $\mathbf{O}$
\end{algorithmic}
\end{algorithm}

\subsection{Recursion}
Obtaining all facets of $\mathcal{F}$, as required in the AD method (see Alg.~\ref{alg:adm}), is again a dual description problem on $\mathcal{F}$, which is of lower dimensionality and has fewer vertices than $\mathcal{P}$. The conversion can be performed by a recursive call of the AD method or any other facet enumeration method. The decision on which method to apply can be based on different heuristics~\cite{Deza2015}, such as number of previous recursive calls, dimensionality or number of vertices.

In this work, the number of vertices of $\mathcal{F}$ is used as a heuristic for recursive applications of the method. If the number of vertices $|\mathcal{V}(\mathcal{F})|$ is smaller than a defined cut-off $n_{cut-off}$, no recursion occurs, but another complete dual description method is applied.

\section{Adjacency Sampling Method}
This section introduces the \textit{Adjacency Sampling (AS) method}, a modification of the AD method, which trades the completeness of the output facets for a drastically decreased runtime (see Alg.~\ref{alg:asm}).

Starting at an initial facet $\mathcal{F}_0$ of $\mathcal{P}$, successive linear programs are applied to descend through sub-faces until reaching a face $\mathcal{K}$, satisfying the heuristic ($|\mathcal{V}(\mathcal{K})| < n_{cut-off}$).
All facets of $\mathcal{K}$ are calculated using a complete dual description method. The rotation algorithm is used to generate the adjacency of $\mathcal{K}$, denoted by $Adj(\mathcal{K})$. The parent polytope $\hat{\mathcal{K}}$ of $\mathcal{K}$ is rotated around the faces in $Adj(\mathcal{K})$, yielding a subset of the adjacency of $\hat{\mathcal{K}}$. This incomplete rotation continues upstream until the facet $\mathcal{F}_0$ is rotated, resulting in new facets $\{ \mathcal{F}_i\} \subset Adj(\mathcal{F}_0)$. In contrast, the general Adjacency Decomposition method would calculate the full adjacency of $\mathcal{F}_0$.

Note that $n_{cut-off}$ controls the tradeoff between completeness and runtime. A higher $n_{cut-off}$ leads to fewer recursive steps, and requires full conversion of a larger sub-polytope. A lower $n_{cut-off}$  decreases runtime since the partial enumeration is typically faster than the complete enumeration.

\begin{algorithm}
\caption{Adjacency Sampling Method}
\label{alg:asm}
\begin{algorithmic}
\State \textbf{Input: } $\mathcal{V}(\mathcal{P})$, Generators of $Aut(\mathcal{P})$, Number of vertices to cut-off $n_{cut-off}$
\State \textbf{Output: } Some facet-classes of $\mathcal{P}$
\State $\mathcal{F}_0 = LP(\mathcal{P})$
\State $\mathbf{C} := \{ \mathcal{F}_0 \}$ \Comment{Facets to be considered}
\State $\mathbf{O} := \{ \mathcal{F}_0 \}$ \Comment{All found facets}
\For{$\mathcal{F} \in \mathbf{C}$}
\While{$|\mathcal{V}(\mathcal{F})| > n_{cut-off}$}
\State $\mathcal{F} = LP(\mathcal{F})$
\EndWhile
\State $\mathbf{H} := \text{All facets of} \; \mathcal{F}$
\While{$\mathcal{F} \neq \mathcal{P}$}
\State $\mathbf{H} = \{ rotate(\mathcal{F}, \mathcal{H}) \, | \, \mathcal{H} \in \mathbf{H} \}$
\State $\mathcal{F} = \text{Parent polytope of} \; \mathcal{F}$
\EndWhile
\For{$\mathcal{H} \in \mathbf{H}$}
\If{$\mathcal{H}$ inequivalent to all facets in $\mathbf{O}$}
\State $\mathbf{C} = \mathbf{C} \cup \{ \mathcal{H} \}$
\State $\mathbf{O} = \mathbf{O} \cup \{ \mathcal{H} \}$
\EndIf
\EndFor
\EndFor
\State \textbf{return} $\mathbf{O}$
\end{algorithmic}
\end{algorithm}

\subsection{Implementation}
In this work, we implement the AS method within the \textit{PANDA} software~\cite{Loerwald2015}. \textit{PANDA} contains a parallelized implementation of the AD method in \textit{C++}. The main additions to the original \textit{PANDA} software are recursive calls and the AS method itself (described in Alg.~\ref{alg:asm}). The equivalence check functionality of \textit{PANDA} is updated with algorithms from \textit{permutalib}~\cite{permutalib}, which re-implements partition-backtracking algorithms of \textit{GAP}~\cite{GAP4}. The implementation is hosted on GitHub~\cite{panda}.

\section{Evaluation}
This section presents the validation of the Adjacency Sampling method and new results from its application to Bell polytopes, that were not (yet) fully enumerated.

\subsection{Validation}
For validation, we use fully enumerated polytopes, comparing the number of found facet-classes to the known number of facet-classes while varying $n_{cut-off}$. The number of found facet-classes for the selected benchmarking polytopes is listed in Tab.~\ref{tab:test_polytopes}. For all selected polytopes, the AS method, even though derived as a sampling method, enumerates all facet-classes previously found in the literature. Using smaller values for $n_{cut-off}$, however, yields an incomplete output.

\begin{table}[htb]
\caption{Benchmarking of the AS method for different local polytopes. The numbers marked with an asterisk are not verified to be complete and are taken from~\cite{Jesus2023}. Results for the cut polytopes are taken from~\cite{Deza2015}. Each enumeration was finished within a few minutes on a single core of an \textit{Intel\textregistered Core\texttrademark i5-1245U}.}
\centering
\begin{tabular}{lrr} 
\toprule
Polytope & \#Facet-Classes & $n_{cut-off}$ \\
\midrule
$\mathcal{L}_{3,3,2,2}$ & 3 / 3 & 20 \\
$\mathcal{L}_{3,2,2,3}$ & 5 / 5 & 20 \\
$\mathcal{L}_{4,3,2,2}$ & 6 / 6 & 30 \\
$\mathcal{L}_{5,3,2,2}$ & 7 / 7 & 40 \\
$\mathcal{L}_{2,2,3,5}$ & 15 / 15 & 20 \\
$\mathcal{L}_{2,2,4,4}$ & 34 / 34* & 60 \\
$\mathcal{L}_{3,2,3,3}$ & 38 / 38 & 40 \\
$\mathcal{L}_{4,3,3,2}$ & 80 / 80* & 90 \\
$\mathcal{L}_{3,3,4,2}$ & 159 / 159* & 70\\
$\mathcal{L}_{4,4,2,2}$ & 175 / 175 & 40\\
\midrule
$K_{4,6}$ & 12 / 12& 10 \\
$K_{1,1,3,3}$ & 50 / 50  & 40 \\
$K_8$ & 147 / 147 & 30 \\
$K_{5,5}$ & 1282 / 1282 & 30 \\
$K_{3,3,3}$ & 2015 / 2015 & 40 \\ 
\bottomrule
\end{tabular}
\label{tab:test_polytopes}
\end{table}

Recently, another sampling method was applied to the polytopes $\mathcal{L}_{2,2,4,4}$, $\mathcal{L}_{3,3,4,2}$ and $\mathcal{L}_{4,3,3,2}$~\cite{Jesus2023}. The AS method recovers exactly the reported facet-class counts (34, 159, 80 classes, respectively). No additional classes were found even after increasing $n_{cut-off}$ up to 140. Since all other polytopes in Tab.~\ref{tab:test_polytopes} were fully enumerated at lower $n_{cut-off}$ values, this provides independent indication that these lists might be complete.

\subsection{New Bell Inequalities}
After validating the method's performance, it is applied to several bipartite Bell polytopes, which have not been fully enumerated previously. All computations were performed on 12 cores of an \textit{Intel\textregistered Xeon\texttrademark E5-1650V3}. The results can be found on Github~\cite{panda}.

\subsubsection*{$\mathcal{L}_{3,3,3,3}$}
This polytope has been addressed with various methods~\cite{Schwarz2016, Cope2019}, leading to the most recent result of 4.8 million classes~\cite{Liu2024}. Using the AS method, we enumerated more than $1.29 \times 10^8$ facet-classes within 5 days, drastically improving the previous result. The method continuously produced new classes until the computation was stopped due to excessive memory usage.

\subsubsection*{$\mathcal{L}_{4,5,2,2}$}
Cope and Colbeck~\cite{Cope2019} reported a partial list of 18\,277 facet-classes. Applying the AS method with $n_{cut-off} = 100$, we identified 49\,358 classes. The method finished within a few minutes, indicating that this list may be complete.

\subsubsection*{$\mathcal{L}_{4,6,2,2}$}
No previous results on this polytope are known to the author. The AS method finished within two days and generated 4\,327\,233 classes using a cut-off of $60$ vertices. Based on the method's performance on smaller polytopes, the list may be complete.

\section{Conclusion}

We have introduced the Adjacency Sampling method, a modification of the Adjacency Decomposition method that enables efficient generation of Bell inequalities for scenarios beyond the reach of exact enumeration techniques. By descending into lower-dimensional sub-faces before rotating, the method explores the facet adjacency structure without requiring full enumeration at each level. The single parameter $n_{cut-off}$ controls the trade-off between coverage and computational cost.

On all tested Bell and Cut polytopes, the method recovers every known facet-class, providing confidence in its effectiveness despite the lack of completeness guarantees. Applied to larger scenarios, it yields substantial numbers of new Bell inequalities: over $1.29 \times 10^8$ classes for $\mathcal{L}_{3,3,3,3}$, 49\,358 classes for $\mathcal{L}_{4,5,2,2}$, and more than 4.3 million classes for $\mathcal{L}_{4,6,2,2}$, the first results for this scenario.

Several directions for future work remain. First, a systematic study of how $n_{cut-off}$ affects completeness could help establish practical criteria for when the output is likely exhaustive. Second, the method could be applied to further Bell scenarios with more parties, settings, or outcomes, as well as to other families of combinatorial polytopes. Finally, the newly generated Bell inequalities may prove useful in applications such as device-independent quantum key distribution, where large sets of tight inequalities can improve the robustness and efficiency of certification protocols.

\normalsize
\bibliography{references_new}

@article{Acin2007,
  author = {Ac\'{\i}n, A. and Brunner, N. and Gisin, N. and Massar, S. and Pironio, S. and Scarani, V.},
  title = {Device-Independent Security of Quantum Cryptography against Collective Attacks},
  journal = {Phys. Rev. Lett.},
  volume = {98},
  pages = {230501},
  year = {2007},
  doi = {10.1103/PhysRevLett.98.230501}
}

@article{Brunner2014,
  author = {Brunner, N. and Cavalcanti, D. and Pironio, S. and Scarani, V. and Wehner, S.},
  title = {Bell nonlocality},
  journal = {Rev. Mod. Phys.},
  volume = {86},
  pages = {419--478},
  year = {2014},
  doi = {10.1103/RevModPhys.86.419}
}

@article{Bell1964,
  author = {Bell, J. S.},
  title = {On the Einstein Podolsky Rosen paradox},
  journal = {Physics Physique Fizika},
  volume = {1},
  pages = {195--200},
  year = {1964},
  doi = {10.1103/PhysicsPhysiqueFizika.1.195}
}

@article{EPR1935,
  author = {Einstein, A. and Podolsky, B. and Rosen, N.},
  title = {Can Quantum-Mechanical Description of Physical Reality Be Considered Complete?},
  journal = {Phys. Rev.},
  volume = {47},
  pages = {777--780},
  year = {1935},
  doi = {10.1103/PhysRev.47.777}
}

@article{Deza2015,
  author = {Deza, M. and Dutour Sikiri\'{c}, M.},
  title = {Enumeration of the facets of cut polytopes over some highly symmetric graphs},
  journal = {Int. Trans. Oper. Res.},
  volume = {23},
  pages = {853--860},
  year = {2015},
  doi = {10.1111/itor.12194}
}

@article{Cope2019,
  author = {Cope, T. and Colbeck, R.},
  title = {Bell inequalities from no-signaling distributions},
  journal = {Phys. Rev. A},
  volume = {100},
  pages = {022114},
  year = {2019},
  doi = {10.1103/PhysRevA.100.022114}
}

@book{Grnbaum2003,
  author = {Gr\"{u}nbaum, B.},
  title = {Convex Polytopes},
  publisher = {Springer},
  year = {2003},
  doi = {10.1007/978-1-4613-0019-9}
}

@article{Christof2001,
  author = {Christof, T. and Reinelt, G.},
  title = {Decomposition and Parallelization Technique for Enumerating the Facets of Combinatorial Polytopes},
  journal = {Int. J. Comput. Geom. Appl.},
  volume = {11},
  pages = {423--437},
  year = {2001},
  doi = {10.1142/S0218195901000560}
}

@article{Loerwald2015,
  author = {L\"{o}rwald, S. and Reinelt, G.},
  title = {{PANDA}: a software for polyhedral transformations},
  journal = {EURO J. Comput. Optim.},
  volume = {3},
  pages = {297--307},
  year = {2015},
  doi = {10.1007/s13675-015-0040-0}
}

@misc{GAP4,
  key = {GAP},
  title = {{GAP} -- {G}roups, {A}lgorithms, and {P}rogramming, Version 4.11.1},
  year = {2021},
  howpublished = {\url{https://www.gap-system.org}}
}

@article{Avis1997,
  author = {Avis, D. and Bremner, D. and Seidel, R.},
  title = {How good are convex hull algorithms?},
  journal = {Comput. Geom.},
  volume = {7},
  pages = {265--301},
  year = {1997},
  doi = {10.1016/S0925-7721(96)00023-5}
}

@book{Ziegler1995,
  author = {Ziegler, G. M.},
  title = {Lectures on Polytopes},
  publisher = {Springer},
  year = {1995},
  doi = {10.1007/978-1-4613-8431-1}
}

@misc{panda,
  author = {Staufenbiel, C.},
  title = {{PANDA} Fork},
  howpublished = {\url{https://github.com/christian512/panda}},
  note = {Accessed: 2026-02-20}
}

@misc{permutalib,
  author = {Dutour Sikiri\'{c}, M.},
  title = {permutalib},
  howpublished = {\url{https://github.com/MathieuDutSik/permutalib}},
  note = {Accessed: 2026-02-20}
}

@book{Deza1997,
  author = {Deza, M. and Laurent, M.},
  title = {Geometry of Cuts and Metrics},
  publisher = {Springer},
  year = {1997},
  doi = {10.1007/978-3-642-04295-9}
}

@inproceedings{Fukuda1996,
  author = {Fukuda, K. and Prodon, A.},
  title = {Double description method revisited},
  booktitle = {Combinatorics and Computer Science},
  publisher = {Springer},
  pages = {91--111},
  year = {1996},
  doi = {10.1007/3-540-61576-8_77}
}

@article{Avis1996,
  author = {Avis, D. and Fukuda, K.},
  title = {Reverse search for enumeration},
  journal = {Discrete Appl. Math.},
  volume = {65},
  pages = {21--46},
  year = {1996},
  doi = {10.1016/0166-218X(95)00026-N}
}

@misc{Joswig2002,
  author = {Joswig, M.},
  title = {Beneath-and-Beyond revisited},
  year = {2002},
  eprint = {math/0210133},
  archivePrefix = {arXiv}
}

@article{Jesus2023,
  author = {Jesus, J. and Zambrini Cruzeiro, E.},
  title = {Tight Bell inequalities from polytope slices},
  journal = {Phys. Rev. A},
  volume = {108},
  pages = {052220},
  year = {2023},
  doi = {10.1103/PhysRevA.108.052220}
}

@article{Hensen2015,
  author = {Hensen, B. and Bernien, H. and Dr\'{e}au, A. E. and Reiserer, A. and Kalb, N. and Blok, M. S. and Ruitenberg, J. and Vermeulen, R. F. L. and Schouten, R. N. and Abell\'{a}n, C. and Amaya, W. and Pruneri, V. and Mitchell, M. W. and Markham, M. and Twitchen, D. J. and Elkouss, D. and Wehner, S. and Taminiau, T. H. and Hanson, R.},
  title = {Loophole-free Bell inequality violation using electron spins separated by 1.3 kilometres},
  journal = {Nature},
  volume = {526},
  pages = {682--686},
  year = {2015},
  doi = {10.1038/nature15759}
}

@article{Zapatero2023,
  author = {Zapatero, V. and van Leent, T. and Arnon-Friedman, R. and Liu, W. and Zhang, Q. and Weinfurter, H. and Curty, M.},
  title = {Advances in device-independent quantum key distribution},
  journal = {npj Quantum Inf.},
  volume = {9},
  pages = {10},
  year = {2023},
  doi = {10.1038/s41534-023-00684-x}
}

@article{Leon1991,
  author = {Leon, J. S.},
  title = {Permutation group algorithms based on partitions, {I}: Theory and algorithms},
  journal = {J. Symb. Comput.},
  volume = {12},
  pages = {533--583},
  year = {1991},
  doi = {10.1016/S0747-7171(08)80103-4}
}

@article{Liu2024,
  author = {Liu, Y. and Chung, H. Y. and Cruzeiro, E. Z. and Gonzales-Ureta, J. R. and Ramanathan, R. and Cabello, A.},
  title = {Equivalence between face nonsignaling correlations, full nonlocality, all-versus-nothing proofs, and pseudotelepathy},
  journal = {Phys. Rev. Res.},
  volume = {6},
  pages = {L042035},
  year = {2024},
  doi = {10.1103/PhysRevResearch.6.L042035}
}

@article{Schwarz2016,
  author = {Schwarz, S. and Bessire, B. and Stefanov, A. and Liang, Y.-C.},
  title = {Bipartite Bell inequalities with three ternary-outcome measurements---from theory to experiments},
  journal = {New J. Phys.},
  volume = {18},
  pages = {035001},
  year = {2016},
  doi = {10.1088/1367-2630/18/3/035001}
}

\end{document}